\documentclass[12pt]{article}

\usepackage{amsmath,amsfonts,latexsym,amssymb,amscd}
\usepackage{pslatex}
\usepackage[latin1]{inputenc}
\usepackage[T1]{fontenc}
\usepackage{pspicture}
\usepackage{verbatim,amsthm,curves,graphics}
\usepackage{mathrsfs}

\usepackage{graphicx}

\newcommand{\G}{{\mathcal G}}
\newcommand{\mr}{{\mathbb R}}

\newcommand{\mc}{{\mathbb C}}

\begin{document}

%\begin{flushright}
%Sofia University\\
%\end{flushright}
%%%%%%%%%%%%%%%%%%%%%%%%%%%%%%%%%%%%%%%%%%%%%%%%%%%%%%%%%%%%%%%%%%%

\title{A classification (uniqueness) theorem for rotating black holes in 4D Einstein-Maxwell-dilaton theory }

\author{
Stoytcho Yazadjiev$^{}$\thanks{\tt yazad@phys.uni-sofia.bg}
\\ \\
{\it $ $Department of Theoretical Physics, Faculty of Physics, Sofia
University} \\
{\it 5 J. Bourchier Blvd., Sofia 1164, Bulgaria} \\
    }
\date{}

\maketitle

\begin{abstract}
In the present paper we prove a classification (uniqueness) theorem for stationary, asymptotically flat  black
hole spacetimes  with connected and non-degenerate horizon in 4D Einstein-Maxwell-dilaton theory with an arbitrary dilaton coupling parameter $\alpha$. We show that such black holes are uniquely specified by
the length of the horizon interval, angular momentum, electric and magnetic charge and the value of the dilaton field at infinity  when the dilaton coupling parameter satisfies $0\le \alpha^2\le3$.
The proof is based on the nonpositivity of the Riemann curvature operator on the space of the potentials. A generalization of the classification theorem for spacetimes with
disconnected horizons is also given.
\end{abstract}

%%%%%%%%%%%%%%%%%%%%%%%%%%%%%%%%%%%%%%%%%%%%%%%%%%%%%%%%%%%%%%%%%%%

%\draft
\sloppy

\section{Introduction}
Key results in the 4D General relativity are the black hole classification (uniqueness) theorems \cite{Israel67}-\cite{Heusler}.
Besides their intrinsic theoretical and mathematical value,  these theorems
have played important role in studying a wide range of topics from astrophysical black holes  to thermodynamics of black holes. The advent of the theoretical physics
leads to gravity theories and models generalizing  general relativity and one of the important problems is the study of  black holes and in particular the classification
of  black hole solutions in  these new theories. An example of such a theory  is the so-called Einstein-Maxwell-dilaton gravity which naturally arises in the context
of the low energy string theory \cite{Gibbons, Garfinkle} and Kaluza-Klein gravity \cite{Maison}. The action for the  Einstein-Maxwell-dilaton gravity is given by
\begin{eqnarray}\label{EMDAction}
{\cal A}= \int d^4x\sqrt{-g} \left(R - 2g^{ab}\partial_{a}\varphi\partial_{b}\varphi - e^{-2\alpha\varphi}F_{ab}F^{ab} \right)
\end{eqnarray}
where $R$ is the Ricci scalar curvature with respect to the spacetime metric $g_{ab}$, $F_{ab}$ is the Maxwell field and $\varphi$ is the dilaton (scalar) field.
The parameter $\alpha$ governs the coupling between the dilaton and electromagnetic field and is called a dilaton coupling parameter.
It is worth mentioning that the Einstein-Maxwell-dilaton gravity arises also in some theories with gradient spacetime  torsion  \cite{Hojman}.

The black holes in 4D Einstein-Maxwell-dilaton theory were extensively studied in various aspects during the last two decades. The static asymptotically flat
Einstein-Maxwell-dilaton black holes were classified in \cite{Masood}. However, no classification is known for the more general case of rotating black holes.
In the present paper we prove  a uniqueness theorem classifying all stationary, asymptotically flat black holes in 4D Einstein-Maxwell-dilaton theory with a certain restriction on
the dilaton coupling parameter.
In proving the classification theorems in pure 4D general relativity one strongly relies on the high degree of symmetries of the dimensionally  reduced stationary and axisymmetric
field equations. In both cases of vacuum Einstein and Einstein-Maxwell equations the space of potentials is a symmetric space which insures the existence of nice properties
and, in particular, insures the existence of the so-called Mazur identity \cite{Mazur},  which is a key point in the uniqueness proof. Contrary to the vacuum Einstein and
Einstein-Maxwell gravity, the space of potentials associated with the dimensionally reduced stationary and axisymmetric Einstein-Maxwell-dilaton   field equations is not a symmetric
space in the general case. This forces us to use in the proof of the classification theorem other intrinsic geometric properties of space of potentials  instead of the absent symmetries
(isometries). Such an intrinsic property is the nonpositivity of the Riemannian curvature operator of the space of potentials  which also insures the nonpositivity of the sectional curvature.
The mentioned property together with the Bunting identity \cite{Bunting},\cite{Carter1} and natural boundary conditions   leads to the proof of desired classification theorem.

The paper is organized as follows. In  Sec. 2 and 3 we give in concise form the necessary mathematical
background. The main result is presented in Sec. 4. In the Discussion we comment on possible  extensions of the classification
theorem and give an explicit extension for spacetimes with disconnected horizons.

\section{Stationary Einstein-Maxwell-dilaton black holes }

Let $(M,g,F,\varphi)$ be a $4$-dimensional, analytic,
stationary black hole spacetime satisfying the Einstein-Maxwel-dilaton equations
\begin{eqnarray}
&&R_{ab}= 2\nabla_a\varphi \nabla_b\varphi  + 2e^{-2\alpha\varphi}  \left(F_{ac}F_{b}{}^c- \frac{g_{ab}}{4}F_{cd}F^{cd} \right),\\
&&\nabla_{a}\left(e^{-2\alpha\varphi}F^{ab}\right)=0=\nabla_{[a} F_{bc]}, \\
&&\nabla_{a}\nabla^{a}\varphi= -\frac{\alpha}{2}e^{-2\alpha\varphi}F_{ab}F^{ab} ,
\end{eqnarray}
derived from the action (\ref{EMDAction}).
Let $\xi$ be the
asymptotically timelike complete Killing field, $\pounds_\xi g = 0$,
which we assume is normalized
so that $\lim g(\xi,\xi)  = -1$ near infinity. We assume also that
the Maxwell tensor and the dilaton field are  invariant under $\xi$, in the sense that $\pounds_\xi F = 0=\pounds_\xi\varphi$.

We consider 4-dimensional spacetime with asymptotic region $M_{\infty}=\mr^{3,1}$ and asymptotic metric

\begin{eqnarray}
g= - dt^2 + dx^2_1 + dx^2_2 + dx^2_3  + O(r^{-1})
\end{eqnarray}
where $x_i$ are the standard Cartesian coordinates on $\mr^3$.
Here $O(r^{-1})$ stands for all metric components that drop off at least as $r^{-1}$ in the radial coordinate $r=\sqrt{x^2_1 + x^2_2 + x^2_3}$.

Denote by $H=\partial B$ the horizon of the black hole $B=M/I^{-}({\cal J}^{+})$ where ${\cal J}^{\pm}$ are the  null infinities of spacetime. As a part of our
technical assumptions we  assume that:

\medskip
\noindent

(i) $H$ is non-degenerate and the horizon cross section is compact  connected manifold of dimension 2.
\medskip
\noindent

(ii) The domain of outer communication $\langle\langle M\rangle \rangle$ is globally hyperbolic.
\medskip
\noindent

According to the topological censorship theorem \cite{Galloway},  $\langle\langle M\rangle\rangle$ is a simply connected manifold with boundary $\partial \langle\langle M \rangle\rangle=H$.
Moreover, one can show that the horizon cross section has spherical topology. {\it In the present paper we shall consider
the case when $\xi$ is not tangent to the null generators of the horizon\footnote{When $\xi$ is tangent to the null generators of the horizon, the spacetime time must be static.
}}. In this case, according to the rigidity theorem \cite{Hawking, Friedrich, HIW} there exists an additional Killing
field $\eta$ which generates a periodic flow  with period $2\pi$, commutes with $\xi$, has nonempty axis  and is such that $\pounds_\eta F=\pounds_\eta\varphi=0$.
In other words, the spacetime is axisymmetric
with an isometry group ${\cal G}=\mr \times U(1)$.   In the asymptotic region $M_{\infty}$ the Killing field $\eta$ takes the standard form

\begin{eqnarray}
&&\eta=x_1{\partial/\partial x_2} -  x_2{\partial/\partial x_1}.
\end{eqnarray}

There exists also  a linear combination

\begin{eqnarray}
{\cal K}= \xi + \Omega_{H} \eta
\end{eqnarray}
so that  the Killing field ${\cal K}$ is tangent and normal to the null generators of the horizon and $g({\cal K},\eta)|_{H}=0$.
The surface gravity $\kappa$ of the black hole  may be defined by ${\cal K}$, namely

\begin{eqnarray}
\kappa = \lim_{H}\frac{\nabla_{a}n\nabla^{a}n}{n}
\end{eqnarray}
where $n=\nabla^a{\cal K}^b\nabla_{a}{\cal K}_{\,\,\,b}$. It is well known that $\kappa$ is constant on the horizon. The horizon non-degeneracy condition
implies  that $\kappa>0$.

Due to the symmetries of the spacetime the natural space to work on is the orbit (factor) space ${\hat M}=\langle\langle M\rangle\rangle/\G$,
where $\G$ is the isometry group. The structure of the factor space in 4D is well known and is described by the following theorem \cite{Carter2}, \cite{Chrusciel},\cite{HY3}:

\medskip
\noindent
{\bf Theorem:} Let $(M\rangle,g)$ be a stationary, asymptotically flat Einstein-Maxwell-dilaton  black hole spacetime with isometry group
$\G=\mr\times U(1)$ satisfying the technical assumptions stated above. Then the orbit space ${\hat M}=\langle \langle M\rangle\rangle /\G$ is
a simply connected 2-dimensional manifold with boundaries and corners homeomorphic to a half-plane. More precisely, the boundary consists of one finite interval $I_{H}$
corresponding to the quotient of the horizon, $I_{H}=H/{\cal G}$ and two semi-infinite intervals $I_{-}$  and  $I_{+}$ corresponding to the axis of $\eta$.
The corners correspond to the points where the axis intersects the horizon.

\medskip
\noindent

In the interior of ${\hat M}$ there is a naturally induced metric ${\hat g}$ which has signature $++$. We denote derivative operator associated with
${\hat g}$ by ${\hat D}$. Let us now consider the Gramm matrix of the Killing fields $\Gamma_{IJ}=g(K_{I},K_{J})$,
where $K_{1}=\xi$  and $K_{2}=\eta$. Then the determinant $\rho^2=-\det \Gamma$ defines a scalar
function $\rho$ on ${\hat M}$ which, as  well known, is harmonic,
${\hat D}^{a}{\hat D}_{a}\rho=0$ as a consequence of the Einstein-Maxwell-dilaton field equations.
It can be shown that $\rho>0$, ${\hat D}_a \rho\ne 0$ in the interior of ${\hat M}$ and that $\rho=0$ on $\partial {\hat M}$.
We may define a conjugate harmonic function $z$ on ${\hat M}$ by $dz={\hat \star}\, d\rho$, where ${\hat \star}$ is the Hodge dual on ${\hat M}$.
The functions $\rho$ and $z$ define global coordinates on ${\hat M}$
identifying the orbit space with the upper complex half-plane
\begin{eqnarray}
{\hat M}= \{z+ i\rho \in \mc, \rho\ge 0  \}
\end{eqnarray}
with the boundary corresponding to the real axis. The induced metric ${\hat g}$ is given in these coordinates by

\begin{eqnarray}
{\hat g}= \Sigma^2(\rho,z)(d\rho^2 + dz^2),
\end{eqnarray}
$\Sigma^2(\rho,z)$ being a conformal factor.

In other words  the orbit space is the half plane ${\hat M}=\{z+i\rho, \rho>0\}$ and its boundary $\partial {\hat M}$ is divided into the intervals
$I_{-}=(-\infty, z_1]$, $I_{H}=[z_1, z_2]$ and $I_{+}=[z_2,+\infty)$:

\begin{eqnarray}
(-\infty,z_1], [z_1,z_2],[z_{2},+\infty) .
\end{eqnarray}
Let us note that the requirement for non-degenerate horizon is equivalent to a non-zero length of the horizon interval, $l(I_{H})=z_2 - z_1>0$.
The parameter $l(I_H)$ is invariantly defined and will play important role in the classification theorem.

\section{Dimensionally reduced Einstein-Maxwell-dilaton  \\equations}

In the context of the present paper it is important to perform the dimensional reduction in terms of the axial Killing field $\eta$. This
insures positively definite metric in the space of the potentials and in this formulation we  also avoid the ergosurface "singularities".
First we consider the twist 1-form $\omega$ associated with $\eta$ and defined by

\begin{eqnarray}
\omega=\star (\eta\wedge d\eta)=i_\eta \star d\eta.
\end{eqnarray}
By definition, $\omega$ is invariant under the symmetries and naturally induces corresponding 1-form ${\hat \omega}$ on the orbit space ${\hat M}$.
Taking  the exterior derivative of $\omega$ we obtain:
\begin{eqnarray}
d\omega= di_\eta \star d\eta =  \left( \pounds_\eta - i_\eta d  \right)\star d\eta = -i_\eta d\star d\eta =
- i_\eta\star d^{\dagger}d\eta = - 2 i_\eta \star R[\eta]
\end{eqnarray}
where $R[\eta]$ is the Ricci 1-form. Making use of the field equations we find

\begin{eqnarray}
d\omega= 4i_{\eta}F\wedge i_{\eta}\left(e^{-2\alpha\varphi}\star F\right).
\end{eqnarray}

As a consequence of the symmetries and the field equations the 1-forms $f=i_\eta F$ and $f_{\star}= i_{\eta}\left(e^{-2\alpha\varphi}\star F\right)$
are closed, i.e. $df=0$ and $df_{\star}=0$. Since  $f$ and $f_{\star}$ are  invariant under the symmetries they naturally induce corresponding 1-form
${\hat f}$ and ${\hat f_{\star}}$ on the orbit space ${\hat M}$ which are still closed. The domain of outer communications   $ \langle\langle M \rangle\rangle$
is simply connected and therefore there exist globally defined potentials $\Phi$ and $\Psi$ such that $f=d\Phi$ and $ f_{\star}=d\Psi$ on $ \langle\langle M \rangle\rangle$.

Proceeding further we have
\begin{eqnarray}
d \omega= 4d\Phi\wedge d\Psi= 2 d\left(\Phi d\Psi - \Psi d\Phi\right).
\end{eqnarray}
Using again the fact that $\langle\langle M\rangle\rangle$ is simply connected we conclude that there exists a potential $\chi$  such that $\omega= d\chi + 2\Phi d\Psi - 2\Psi d\Phi$
on $\langle\langle M\rangle\rangle$. Obviously, the potentials $\chi$, $\Phi$ and $\Psi$ are invariant under the symmetries and they are naturally defined on the orbit space ${\hat M}$.

The potential $\Phi$, $\Psi$ and $\chi$ play important role in writing down the dimensionally reduced Einstein-Maxwell-dilaton equations on the orbit space. Let $X$ and $h$
be functions on ${\hat M}$ defined by

\begin{eqnarray}
X=g(\eta,\eta), \,\,\,\,\,\, e^{2h} X^{-1}= g(d\rho,d\rho).
\end{eqnarray}

Then the stationary and axisymmetric Einstein-Maxwell-dilaton  equations are equivalent to the following set of
equations on the orbit space $\hat M$ :

\begin{eqnarray}
&&\rho^{-1} {\hat D}_a\left(\rho {\hat D}^a X\right) =X^{-1} {\hat D}_a X {\hat D}^a X -   X^{-1}\left({\hat D}_a\chi + 2\Phi {\hat D}_a\Psi - 2\Psi {\hat D}_a\Phi\right) \left({\hat D}^a\chi
+ 2\Phi {\hat D}^a\Psi - 2\Psi {\hat D}^a\Phi\right)  \nonumber \\
&& -2 e^{-2\alpha\varphi} {\hat D}_a\Phi {\hat D}^a\Phi  -2 e^{2\alpha\varphi} {\hat D}_a\Psi {\hat D}^a\Psi ,
 \label{dreq1}\\ \nonumber  \\
&&{\hat D}_a \left[\rho X^{-2} \left({\hat D}^a\chi + 2\Phi {\hat D}^a\Psi - 2\Psi {\hat D}^a\Phi \right)\right]=0, \\ \nonumber \\
&& \rho^{-1}{\hat D}_a\left( X^{-1}\rho e^{-2\alpha\varphi} {\hat D}^a\Phi\right)= X^{-2} {\hat D}_a\Psi \left( {\hat D}^a \chi
+ 2\Phi {\hat D}^a\Psi - 2\Psi {\hat D}^a\Phi\right), \\ \nonumber \\
&&\rho^{-1}{\hat D}_a\left( X^{-1}\rho e^{2\alpha\varphi} {\hat D}^a\Psi\right)= - X^{-2} {\hat D}_a\Phi \left( {\hat D}^a \chi
+ 2\Phi {\hat D}^a\Psi - 2\Psi {\hat D}^a\Phi\right), \\ \nonumber \\
&&\rho^{-1} {\hat D}_{a}\left( \rho {\hat D}^{a}\varphi\right)= -\alpha\, X^{-1}\left( e^{-2\alpha\varphi} {\hat D}_a\Phi {\hat D}^{a}\Phi  -
e^{2\alpha\varphi} {\hat D}_a\Psi {\hat D}^{a}\Psi  \right), \label{dreqlast}
\end{eqnarray}
together with

\begin{eqnarray}
&&\rho^{-1}{\hat D}^a \rho {\hat D}_a h = \left[\frac{X^{-2}}{4} {\hat D}^a X {\hat D}^bX
+ \frac{X^{-2}}{4}  \left({\hat D}^a\chi + 2\Phi {\hat D}^a\Psi - 2\Psi {\hat D}^a\Phi\right) \left({\hat D}^b\chi + 2\Phi {\hat D}^b\Psi - 2\Psi {\hat D}^b\Phi\right)   \right. \nonumber\\
&&\left. +   X^{-1} e^{-2\alpha\varphi} {\hat D}^a\Phi {\hat D}^b\Phi  + X^{-1}e^{2\alpha\varphi} {\hat D}^a\Psi {\hat D}^b\Psi + {\hat D}^a\varphi {\hat D}^b\varphi \right]
\cdot \left[ {\hat g}_{ab} - 2 {\hat D}_a z {\hat D}_b z\right] ,\label{heq1} \\ \nonumber \\
&&\rho^{-1} {\hat D}^a z {\hat D}_a h = 2  \left[\frac{X^{-2}}{4} {\hat D}^a X {\hat D}^bX
+ \frac{X^{-2}}{4}  \left({\hat D}^a\chi + 2\Phi {\hat D}^a\Psi - 2\Psi {\hat D}^a\Phi\right) \left({\hat D}^b\chi + 2\Phi {\hat D}^b\Psi - 2\Psi {\hat D}^b\Phi\right)   \right.
\nonumber\\
&&\left. +   X^{-1} e^{-2\alpha\varphi} {\hat D}^a\Phi {\hat D}^b\Phi
+ X^{-1}e^{2\alpha\varphi} {\hat D}^a\Psi {\hat D}^b\Psi + {\hat D}^a\varphi {\hat D}^b\varphi \right] {\hat D}_a\rho {\hat D}_b z  .\label{heq2}
\end{eqnarray}

The equations (\ref{heq1}) and (\ref{heq2}) are decoupled from the group
 equations (\ref{dreq1}) - (\ref{dreqlast}).  Once the solution of the system equations (\ref{dreq1}) - (\ref{dreqlast}) is known we can  determine the function $h$.
 Therefore the problem of the classification  of the 4D Einstein-Maxwell-dilaton black holes can be studied as 2-dimensional boundary value problem for the nonlinear
 partially differential equation system  (\ref{dreq1}) - (\ref{dreqlast}) as the boundary conditions are specified below.

 At this stage we introduce the strictly positive\footnote{The axial Killing field $\eta$ is strictly spacelike
everywhere, i.e.  $X=g(\eta,\eta)>0$ except on the symmetry axis where $\eta$ vanishes. In the present work we exclude the possibility of closed timelike curves. } definite metric

 \begin{eqnarray}\label{harmetric}
 dL^2 = G_{AB}dX^A dX^B= \frac{dX^2 + \left(d\chi + 2\Phi d\Psi - 2\Psi d\Phi \right)^2 }{4X^2} + \frac {e^{-2\alpha\varphi}d\Phi^2
 + e^{2\alpha\varphi} d\Psi^2}{X} + d\varphi^2
 \end{eqnarray}
on the 5-dimensional manifold ${\cal N}=\{(X,\chi, \Phi, \Psi,\varphi) \in \mr^5; X>0\}$.

The equations (\ref{dreq1}) - (\ref{dreqlast}) can be obtained from a variational principle based on the functional
\begin{eqnarray}\label{functional}
I[X^A]=\int_{{\hat M}} d^2x \,\rho \sqrt{{\hat g}}\, {\hat g}^{ab} G_{AB}(X^C) \partial_{a}X^A \partial_{b} X^B .
\end{eqnarray}

Further we consider the mapping
\begin{eqnarray}\label{harmonicmapp}
 {\cal X}: {\hat M} \mapsto {\cal N}
\end{eqnarray}
of the 2-dimensional Riemannian manifold  ${\hat  M}$ onto the 5-dimensional Riemannian manifold ${\cal N}$ the local  coordinate representation
of which
\begin{eqnarray}
{\cal X}: (\rho,z) \mapsto X^{A}
\end{eqnarray}
satisfies the equations  (\ref{dreq1}) - (\ref{dreqlast}) derived from the functional (\ref{functional}). It is well known that ${\cal X}$ belongs
to the class of the so-called harmonic maps.

We shall close this section with important comments. The  Riemannian manifold $({\cal N}, G_{AB})$ is not a symmetric space in the general case
for arbitrary dilaton coupling parameter $\alpha$. This is seen from the fact that for the Riemannian curvature tensor we have
\begin{eqnarray}
\nabla_{E}R_{ABCD}\ne 0
\end{eqnarray}
as one can check. Only in the cases $\alpha=0$ and $\alpha^2=3$,  $({\cal N},G_{AB})$ is a symmetric space.
The first case corresponds to the pure Einstein-Maxwell gravity for which it is well known that metric $G_{AB}$ possesses the maximal group of symmetries (isometries)
which in turn insures the complete integrability of the stationary and axi-symmetric Einstein-Maxwell equations.
The second case corresponding to 4D Kaluza-Klein gravity  inherits its symmetries via the dimensional reduction of the 5D vacuum Einstein gravity
which also possesses high degree of symmetry, especially for $\mr \times U(1)^2$ group of spacetime isometries.

\section{Classification theorem }

We start this section with intermediate results which will be used in the central classification theorem.

\medskip
\noindent
{\bf Lemma:} {\it The manifold $({\cal N}, G_{AB})$ is geodesically complete for any $\alpha$.}

\medskip
\noindent

{\bf Proof:} Let $s\mapsto \gamma(s)$ be an affinely parameterized geodesic,

\begin{eqnarray}
\gamma(s)=\left(X(s),\chi(s),\Phi(s),\Psi(s),\varphi(s)\right) .
\end{eqnarray}

Then $G(\dot \gamma,\dot \gamma)=C>0$ is a constant of motion, i.e.

\begin{eqnarray}
G_{AB}\dot X^A\dot X^B= \frac{1}{4} \left(\frac{\dot X}{X}\right)^2 + \frac{\left(\dot\chi + 2\Phi \dot\Psi - 2\Psi \dot\Phi \right)^2}{4X^2}
+ \frac {e^{-2\alpha\varphi}\dot\Phi^2}{X}  + \frac {e^{2\alpha\varphi} \dot\Psi^2}{X} + \dot\varphi ^2 = C.
\end{eqnarray}

Hence it follows that we have

\begin{eqnarray}
\frac{1}{4} \left(\frac{\dot X}{X}\right)^2 \le C , \;\;\;\;\;\;\;\;\;  \dot \varphi^2 \le C ,
\end{eqnarray}

and
\begin{eqnarray}
\dot \Phi^2 \le C X e^{2\alpha\varphi},
\;\;\;\;\;\;\; \;\; \dot \Psi^2 \le C X e^{-2\alpha\varphi}, \;\;\; \;\;\;\; \left(\dot\chi + 2\Phi \dot\Psi - 2\Psi \dot\Phi \right)^2\le 4C X^2 .
\end{eqnarray}
Therefore any geodesic can be extended to arbitrary values of the affine parameter, i.e. the metric is geodesically complete. Even though there is an edge at $X=0$ this
edge is at an infinite distance from any point of the manifold.

\medskip
\noindent

{\bf Lemma:} {\it The manifold $({\cal N}, G_{AB})$ is manifold with nonpositive Riemann curvature operator
for a dilaton coupling parameter  $\alpha$ satisfying  $0\le \alpha^2\le 3$.}

\medskip
\noindent

\medskip
\noindent

{\bf Proof:} Let $\Lambda^2T^{*}({\cal N})$ be the linear space of 2-forms on ${\cal N}$. We regard the Riemann curvature tensor  $Riemann =\{R_{ABCD}\}$
as an operator
\begin{eqnarray}
 {\cal \hat R}\; : \Lambda^2T^{*}({\cal N}) \rightarrow \Lambda^2T^{*}({\cal N}).
\end{eqnarray}
which is symmetric with respect  to the naturally induced metric\footnote{For separable forms  $x\wedge y$ and $u\wedge w$  the metric $\langle,\rangle$ is defined by
$\langle x\wedge y, u\wedge w \rangle=G(x,u)G(y,w) - G(x,w)G(y,u)$. }   $\langle,\rangle$  on $\Lambda^2T^{*}({\cal N})$. The nonpositivity of  the curvature
operator then  means that  $\langle {\cal \hat R}\Omega, \Omega\rangle=\langle \Omega, {\cal \hat R}\Omega\rangle= {\cal R}(\Omega,\Omega)\le0$ for all $\Omega \in \Lambda^2T^{*}({\cal N})$.

In order to show the non-positivity of the Riemannian operator ${\cal \hat R}$ we fix an orthonormal basis $\{\omega^{\,i}\wedge \omega^{j}\}$  in $\Lambda^2T^{*}({\cal N})$
where $\{\omega^{\,i}\}$ is an orthonormal  basis in  $T^{*}{\cal N}$ explicitly given by
\begin{eqnarray}
&& \omega^{1}=\omega^{X}= \frac{dX}{2X}, \,\, \,\,\, \,\,\,\,\,\,\,\,\,\,\,\,\,\omega^{2}=\omega^{\chi}= \frac{d\chi + 2 \Phi d\Psi - 2\Psi d\Phi}{2X} ,  \nonumber \\ \\
&&\omega^{3}=\omega^{\Phi}=\frac{e^{-2\alpha\varphi}d\Phi}{\sqrt{X}}, \,\,\,\,\,\,\,\,\,\,\,\,\,\,\, \omega^{4}=\omega^{\Psi}= \frac{e^{2\alpha\varphi}d\Psi}{\sqrt{X}},
\,\,\,\,\,\,\,\, \,\,\, \,\,\,\,\,\,\,\,\omega^{5}=\omega^{\varphi}= d\varphi . \nonumber
\end{eqnarray}

 In this orthonormal basis the matrix representing the Riemann operator is symmetric and is explicitly given by

\begin{eqnarray}
{\cal \hat R}=
         \left(
           \begin{array}{cccccccccc}
             -4 & 0 & 0 & 0 & 0 & 0 & 0 & 2 & 0 & 0 \\
             0 & -1 & 0 & 0 & 0 & 1 & 0 & 0 &\alpha  & 0 \\
             0 & 0 & -1 & 0 & -1 & 0 & 0 & 0 & 0 & -\alpha \\
             0 & 0 & 0 & 0 & 0 & 0 & 0 & 0 & 0 & 0 \\
             0 & 0 & -1 & 0 & -1 & 0 & 0 & 0 & 0 & -\alpha \\
             0 & 1 & 0 & 0 & 0 & -1 & 0 & 0 & -\alpha & 0 \\
             0 & 0 & 0 & 0 & 0 & 0 & 0 & 0 & 0 & 0 \\
             2 & 0 & 0 & 0 & 0 & 0 & 0 & \alpha^2-4 & 0 & 0 \\
             0 & \alpha & 0 & 0 & 0 & -\alpha & 0 & 0 & -\alpha^2 & 0 \\
             0 &  0& -\alpha & 0 & -\alpha & 0 & 0 & 0 & 0 & -\alpha^2 \\
           \end{array}
         \right).
\end{eqnarray}

The eigenvalues of this symmetric matrix are

\begin{eqnarray}
&&\lambda_1= - 4 + \frac{\alpha^2}{2} + \frac{\sqrt{16 + \alpha^4}}{2}, \;\;\; \;\;\;\;\; \lambda_2= - 4 + \frac{\alpha^2}{2} - \frac{\sqrt{16 + \alpha^4}}{2},
\\\nonumber  \\
&&\lambda_3 =  \lambda_4= -2- \alpha^2, \;\;\;\;\;\;\;\;\;\;  \lambda_5=\lambda_6=\lambda_7=\lambda_8=\lambda_9=\lambda_{10}=0 .\nonumber
\end{eqnarray}
Obviously, for $0\le \alpha^2\le 3$ all the eigenvalues are nonpositive and therefore the Riemann operator is nonpositive. As an immediate consequence we
also have that ${\cal N}$ is a manifold with nonpositive sectional curvature for $0\le \alpha^2\le 3$ .

\medskip
\noindent

Before going to the formulation of the classification theorem let us give the definitions of the angular momentum $J$, electric charge $Q$ and magnetic charge $P$.
They are respectively given  by the integrals

\begin{eqnarray}
&&J= \frac{1}{16\pi} \int_{S^2_{\infty}} \star d\eta ,  \nonumber \\ \nonumber \\
&&Q= \frac{1}{4\pi} \int_{S^2_{\infty}} e^{-2\alpha\varphi}\star F, \\\nonumber \\
&&P= \frac{1}{4\pi} \int_{S^2_{\infty}}  F, \nonumber
\end{eqnarray}
where $S^2_{\infty}$ is the sphere at infinity.

\medskip
\noindent

\medskip
\noindent
{\bf Classification (Uniqueness) Theorem:} {\it There can be at most only one stationary, asymptotically flat black hole spacetime satisfying the Einstein-Maxwell-dilaton field equations
and the technical assumptions stated  in Sec. 2 for a given set of parameters $\{l(I_H), J, Q, P\}$ and for a dilaton coupling parameter $\alpha$ satisfying $0\le \alpha^2\le 3$.}

\medskip
\noindent

{\bf Remark:} In the theorem we assume that $\lim_{\infty}\varphi = \varphi_\infty=0$. When $\varphi_\infty\ne 0$ it is easy to see that the
set of parameters should be expanded to $\{l(I_H), J, Q, P, \varphi_{\infty}\}$

\medskip
\noindent
{\bf Proof:} Consider two solutions $(M, g, F, \varphi)$ and $({\tilde M}, {\tilde g}, {\tilde F}, {\tilde \varphi})$ as in the statement of the theorem.
We use the same "tilde" notation to distinguish any quantities associated with the two solutions.
Since $l(I_H)=l({\tilde I_H})$,  we can identify the orbit spaces ${\hat M}$ and ${\hat {\tilde M}}$. Moreover we can identify $\langle\langle M\rangle\rangle$ and
$\langle\langle{\tilde M}\rangle\rangle$ as manifolds with $\mr \times U(1)$-action since they can be uniquely reconstructed from the orbit space. We may therefore
assume that $\langle\langle M\rangle\rangle=\langle\langle{\tilde M}\rangle\rangle$ and that $\xi={\tilde \xi}$, $\eta={\tilde \eta}$. We may also assume that
$\rho={\tilde \rho}$  and $z={\tilde z}$.

As a consequence of these identifications, $(g, F, \varphi)$  and $({\tilde g}, {\tilde F}, {\tilde \varphi})$  may be considered as being defined on the same
manifold.  Moreover, on the base of these identifications, we can combine the two solutions into a single identity playing a key role in the proof  as we will see below.

Further we consider two harmonics maps  (\ref{harmonicmapp}) ${\cal X}: {\hat M} \mapsto {\cal N}$ and ${\tilde {\cal X}}: {\hat M} \mapsto {\cal N}$ and
a smooth  homotopy

\begin{eqnarray}
{\cal T}: {\hat M}\times [0,1] \mapsto {\cal N}
\end{eqnarray}
 so that
\begin{eqnarray}
{\cal T}(\tau=0)={\cal X}, \;\;\;\; \;{\cal T}(\tau=1)={\tilde {\cal X}}
\end{eqnarray}
where  $0\le \tau\le 1$ is the homotopy parameter. Rephrasing in local terms we consider  two solutions $X^{A}(\rho,z)$ and ${\tilde X}^A(\rho,z)$  and
a smooth homotopy ${\cal T}: {\hat M}\times [0,1] \mapsto {\cal N}$ such that
\begin{eqnarray}
{\cal T}^{A}(\rho,z;\tau=0)=X^A(\rho,z) ,\;\;\;\;\;\;   {\cal T}^{A}(\rho,z;\tau=1)={\tilde X}^{A}(\rho,z)
\end{eqnarray}
for each point $(\rho,z)\in {\hat M}$.    As a further requirement we impose that the
curves $[0,1]\mapsto {\cal N}$  be geodesic which in local coordinates means that

\begin{eqnarray}
\frac{dS^A}{d\tau} + \Gamma^{A}_{B\,C}S^{B}S^{C}=0
\end{eqnarray}
where
\begin{eqnarray}
S^A=\frac{d{\cal T}^A}{d\tau}
\end{eqnarray}
is the tangent vector along the curves and $\Gamma^{A}_{B\,C}$ are components of the Levi-Civita connection on ${\cal N}$.

The existence and uniqueness of the geodesic homotopy follow from the Hadamard-Cartan theorem \cite{Postnikov} since the Riemannian  manifold $({\cal N}, G_{AB})$
is geodesically complete, simply connected and with nonpositive sectional curvature.

The length of the geodesics will be denoted by $S$, i.e.

\begin{eqnarray}
S= \int^{1}_{0} d\tau\, G_{AB}S^A S^B .
\end{eqnarray}
Let us also note that the tangent vector satisfies the $\tau$-independent normalization condition

\begin{eqnarray}
S^A S_A=S^2.
\end{eqnarray}

Now we are at position to write the  Bunting identity \cite{Bunting},\cite{Carter1}

\begin{eqnarray}\label{Bidentity}
{\hat D}_a\left(\rho S{\hat D}^a S \right)= \rho \int^{1}_{0} d\tau \left({\hat \nabla}^a S_A {\hat \nabla }_a S^A
 - R_{ABCD} S^A {\hat \nabla}_a{\cal T}^B  S^C {\hat \nabla}^a {\cal T}^D  \right).
\end{eqnarray}
Here ${\hat \nabla}_a$ is  the induced connection along the harmonic map, i.e.

\begin{eqnarray}
{\hat \nabla}_a V^A  = \partial_a{\cal T}^C \nabla_{C} V^A= \partial_a V^A +    \partial_a {\cal T}^C \Gamma^{A}_{CB} V^{B}.
\end{eqnarray}

Since $\rho\ge 0$, ${\cal N}$ has positive definite metric and nonpositive sectional curvature for $0\le \alpha^2\le 3$ we conclude that

\begin{eqnarray}
{\hat D}_a\left(\rho S{\hat D}^a S \right)\ge 0 .
\end{eqnarray}

Applying the Gauss theorem we obtain

\begin{eqnarray}\label{Gauss}
\int_{{\hat M}}{\hat D}_a\left(\rho S{\hat D}^a S \right)\sqrt{{\hat g}}d^2x  = \int_{\partial {\hat M}\cup \infty}\rho S {\hat D}^a S d\Sigma_a.
\end{eqnarray}

We shall show that the boundary integral on the right hand side of  (\ref{Gauss}) vanishes. To do this we have to consider the boundary conditions.
In the asymptotic region ($r\to \infty$) we have the standard boundary conditions

\begin{eqnarray}\label{conditions_infinity}
&&X=\rho^2\left(1 + {\cal O}(r^{-1})\right)=r^2\sin^2\theta\left(1 + {\cal O}(r^{-1})\right), \nonumber  \\ \nonumber \\
&&\chi = 2J\cos\theta (2+ \sin^2\theta) + {\cal O}(r^{-1}), \nonumber\\ \nonumber \\
&&\Phi = P\cos\theta + {\cal O}(r^{-1}), \\  \nonumber\\
&&\Psi = Q\cos\theta + {\cal O}(r^{-1}), \nonumber \\\nonumber \\
&&\varphi =  {\cal O}(r^{-1}), \nonumber
\end{eqnarray}
where we have introduced the asymptotic coordinates $r$ and $\theta$ given by $\rho=r\sin\theta$ and $z=r\cos\theta$.

On the horizon interval  the fields  have to be regular and we have remarkably simple boundary conditions

\begin{eqnarray}\label{conditions_horizon}
X= {\cal O}(1),\;\;\; \chi = {\cal O}(1), \;\;\; \Phi={\cal O}(1),\;\;\; \Psi={\cal O}(1), \;\;\; \varphi={\cal O}(1) .
\end{eqnarray}

The regularity on the axis intervals $I_{\pm}$   requires the following boundary conditions

\begin{eqnarray}\label{conditions_axis}
&&X={\cal O}(\rho^2), \nonumber \\ \nonumber \\
&&\chi = \pm 4J+ {\cal O}(\rho^2) ,\nonumber\\ \nonumber \\
&&\Phi= \pm P + {\cal O}(\rho^2) ,\\ \nonumber \\
&&\Psi = \pm Q + {\cal O}(\rho^2), \nonumber \\ \nonumber\\
&&\varphi =   {\cal O}(1). \nonumber
\end{eqnarray}

These boundary conditions can be derived as  follows. Since  $i_\eta F$ and $i_{\eta}e^{-2\alpha\varphi}\star F$ vanish on the axis   the potentials $\Phi$  and $\Psi$ are constant
on the semi-infinite  intervals $I_{-}$ and $I_{+}$ and comparing with the asymptotic behavior (\ref{conditions_infinity}) we conclude that constants are exactly $\pm P$ and $\pm Q$.
Together with the fact that $d\Phi=i_\eta F$ and $d\Psi=i_{\eta}e^{-2\alpha\varphi}\star F$ also vanish on the axis we conclude that near the axis we have
$\Phi=\pm P + {\cal O}(\rho^2)$ and $\Psi= \pm Q + {\cal O}(\rho^2)$. The same considerations can be applied to the twist  potential $\chi$. By definition $\omega$
vanishes on the axis which together with the fact that $d\Phi=d\Psi=0$ on the axis, implies that $\chi$ is a constant on the axis. From the asymptotic behavior we find that
constant is exactly $\pm 4J$ which gives $\chi=\pm 4J + {\cal O}(\rho^2)$.

It is worth noting that the boundary conditions are formally independent of the dilaton coupling constant $\alpha$ and have the same form as in the pure Einstein-Maxwell gravity.

Let us go back to the boundary integral on the right hand side of  (\ref{Gauss}). The contribution from the boundary intervals corresponding to the  horizon and the symmetry axis  in $\partial {\hat M}$ vanishes due to the boundary conditions
(\ref{conditions_horizon}) and (\ref{conditions_axis}) and the fact that $\rho=0$ on the mentioned intervals.
The asymptotic boundary conditions  imply vanishing of the line integral on the infinite arc where $\rho\to \infty$.
Therefore we have
\begin{eqnarray}
\int_{{\hat M}}{\hat D}_a\left(\rho S{\hat D}^a S \right)\sqrt{\hat g}d^2x  =0
\end{eqnarray}
which means that $S$ is constant everywhere.
Since $S$ vanishes at infinity, so it must vanish everywhere. This completes the proof.

Let us note that the black hole uniqueness theorem for the pure 4D Einstein-Maxwell equations follows from our classification theorem as  a particular case for $\alpha=0$.

As a consequence of the theorem one might expect that there would be non-unique black hole solutions in Einstein-Maxwell-dilaton gravity with dilaton coupling parameter
$\alpha^2>3$. Signs for the existence of such non-uniqueness  seem to be found numerically in \cite{Kunz}. At least, the results of \cite{Kunz} show that
beyond the critical value $\alpha=\sqrt{3}$ there are strange and interesting stationary Einstein-Maxwell-dilaton black hole solutions.

\section{Discussion}

In this paper we proved  a classification theorem for stationary rotating asymptotically flat black holes in 4D Einstein-Maxwell-dilaton theory.
The present results can be extended in several directions. The extension of the classification theorem to the case with a degenerate horizon seems to be more delicate
and the formal proof will be presented elsewhere. The generalization of the theorem for spacetimes with disconnected non-degenerate horizons is as follows.
The orbit space for spacetimes with disconnected horizons is again the upper half plane with boundary divided by intervals in the form
\begin{eqnarray}
(-\infty, z_1][z_1,z_2][z_2,z_3].......[z_{(2N-1)},z_{2N}][z_{2N},+\infty)
\end{eqnarray}
 with $N$ finite intervals corresponding to the factor spaces of the horizons and  $N-1$ finite intervals corresponding to the axes between the horizons.
 Finite intervals associated with the horizons and axes between the horizons we denote by $I^{H}_i$ and $I^{axis}_j$, respectively.  The semi-infinite axis
 intervals are again denoted  by $I_{\pm}$. The described structure of the boundary of the factor space together with the lengths of the finite intervals
 $l(I^{H}_i)$ and $l(I^{axis}_j)$ we call {\it interval structure} \cite{HY3}.

The boundary conditions at infinity are the same as (\ref{conditions_infinity}) with the only difference that $J$, $Q$ and $P$ are the total angular momentum,
the total electric and the total magnetic charge, i.e. $J=\sum_i {\tilde J}^{\,H}_{i}$, $Q=\sum_i Q^{H}_i$ and $P=\sum_i P^{H}_i$. Here ${\tilde J}^{\,H}_{i}$
is the total angular momentum of the $i$-th horizon, i.e. the angular momentum with the electromagnetic field contribution taken into account

\begin{eqnarray}
{\tilde J}^{\,H}_{i}= J^{H}_{i} - \frac{1}{4} \int_{I^{H}_i} \left( \Phi d\Psi -\Psi d\Phi \right)= \frac{1}{8}\int_{I^{H}_i} d\chi
\end{eqnarray}
where

\begin{eqnarray}
J^{H}_{i}= \frac{1}{16\pi}\int_{H_i} \star d\eta = \frac{1}{8} \int_{I^{H}_i} \omega.
\end{eqnarray}

On the intervals $I^{H}_i$ corresponding to the horizons the boundary conditions  are the same as (\ref{conditions_horizon}).
On the intervals $I^{axis}_{i}$ we have the following boundary conditions:

\begin{eqnarray}
&&X={\cal O}(\rho^2), \nonumber \\ \nonumber \\
&&\chi= 4\sum^{i}_{k=1} {\tilde J}^{\,H}_k - 4\sum^{N}_{k=i+1} {\tilde J}^{\,H}_k + {\cal O}(\rho^2)=-4J + 8 \sum^{i}_{k=1} {\tilde J}^{\,H}_k + {\cal O}(\rho^2) , \nonumber \\ \nonumber \\
&&\Phi= \sum^{i}_{k=1} P^{H}_{k} - \sum^{N}_{k=i+1} P^{H}_{k} + {\cal O}(\rho^2)= -P + 2 \sum^{i}_{k=1} P^{H}_k + {\cal O}(\rho^2), \\ \nonumber\\
&&\Psi= \sum^{i}_{k=1} Q^{H}_{k} - \sum^{N}_{k=i+1} Q^{H}_{k} + {\cal O}(\rho^2)=-Q +  2\sum^{i}_{k=1} Q^{H}_k + {\cal O}(\rho^2),\nonumber \\ \nonumber\\
&& \varphi = {\cal O}(1) .\nonumber
\end{eqnarray}
On the semi-infinite intervals $I_{\pm}$ the boundary conditions are as  before $\chi=\pm 4J + {\cal O}(\rho^2)$, $\Phi=\pm P + {\cal O}(\rho^2)$,
$\Psi=\pm Q + {\cal O}(\rho^2)$,  $\varphi= O(1)$.

Again the boundary conditions are formally independent of the dilaton coupling parameter $\alpha$ and have the same form as for the pure Einstein-Maxwell gravity. Having the boundary conditions and repeating the same steps as in the previous section one can prove the following theorem.

\medskip
\noindent

{\bf Theorem:} {\it There can be at most only one stationary, axisymmetric and  asymptotically flat black hole spacetime with non-degenerate horizons  satisfying the Einstein-Maxwell-dilaton equations for a given interval structure, given set of parameters $\{{\tilde J}^{\,H}_{i}, Q^{H}_i,P^{H}_i \}$ associated with the horizons, given $\varphi_{\infty}$  and for dilaton coupling parameter satisfying $0\le \alpha^2\le 3$.}

\medskip
\noindent

It is important to note that, in the general case, the  spacetimes with disconnected horizons will have conical singularities on the axes between the horizons.
The question whether there can be completely regular solutions without conical singularities is at present unclear.

It seems also possible to extend the classification theorem to the case of rotating asymptotically non-flat Einstein-Maxwell-dilaton black holes. In particular, the
uniqueness theorem could be extended  for rotating Einstein-Maxwell-dilaton black hole solutions with $C$-metric  and Melvin-Ernst asymptotic generalizing in this way
the uniqueness theorems for static $\alpha=1$ Einstein-Maxwell-dilaton black holes with $C$-metric  and Melvin-Ernst asymptotic  \cite{Wells}.

An important step  will be  the generalization  of our results to the case of higher dimensional spacetimes. In higher dimensions and  in five dimensions in particular,
even the potential space of the  pure Einstein-Maxwell gravity is not a symmetric space. That is why the approach of present paper will be crucially important for
proving the general uniqueness theorems for rotating Einstein-Maxwell-dilaton black holes in both cases of asymptotically flat  and Kaluza-Klein spacetimes.
In other words using the present approach we would be able to generalize the  uniqueness theorems of \cite{HY2} and \cite{Y} for the case when the electromagnetic
field  is fully excited.

\vskip 1cm

\noindent
{\bf Acknowledgements:}
This work was partially supported by the Bulgarian National Science Fund under Grants DO 02-257, VUF-201/06 and by Sofia University Research Fund
under Grant No 101/2010.

\end{document}